\documentclass[prl,twocolumn,nofootinbib,showpacs,superscriptaddress,amsmath,amssymb,amsfonts]{revtex4}
\usepackage{bm}
\usepackage{color,xcolor}
\usepackage{mathtools,amsbsy}
\usepackage{keyval,graphicx}
\usepackage{slashed}
\usepackage{epstopdf}
\usepackage{isotope}
\usepackage{dcolumn}
\def\vec#1{\boldsymbol{#1}}
\begin{document}
\title{Understanding the shortened lifetime of \boldmath$\isotope[3][\Lambda]{H}$\unboldmath}

\author{Jean-Marc Richard}\email{j-m.richard@ipnl.in2p3.fr}
\affiliation{Universit\'e de Lyon, Institut de Physique Nucl\'eaire, UCBL--IN2P3-CNRS,
4, rue Enrico Fermi, Villeurbanne, France}
\author{Qian Wang}\email{wangqian@hiskp.uni-bonn.de}
\affiliation{Helmholtz-Institut f\"ur Strahlen- und Kernphysik and Bethe
Center for Theoretical Physics, \\Universit\"at Bonn,  D-53115 Bonn, Germany}
\author{Qiang Zhao}\email{zhaoq@ihep.ac.cn}
\affiliation{Institute of High Energy Physics and Theoretical Physics Center for Science Facilities,\\
        Chinese Academy of Sciences, Beijing 100049, China}

\begin{abstract}

The lifetime of the hypertriton $\isotope[3][\Lambda]{H}$ has been recently measured as significantly shorter than that of the free $\Lambda$. We present an explanation based on a change of the intrinsic hyperon decay due to the nuclear environment.
\end{abstract}
\date{\today}
\maketitle

The study of hypernuclei is an important tool for probing the hyperon-nucleon interactions and gaining insights into the strong interaction dynamics that binds the hyperons and nucleons to form novel nuclear states.

The hadronic weak decay of hyperons involves two typical processes: (i) the direct weak emission of the pion from the $s$ quark via a four-quark interaction, and (ii) the baryon internal conversion by the weak interaction in association with a strong pion emission. As studied in the literature \cite{LeYaouanc:1988fx}, the process (ii), though model dependent, is much larger than  (i) in the hadronic weak decays of $\Lambda$ and $\Sigma$.

During the past decades, the experimental study of hypernuclei has been significantly improved.
In particular, Rappold {\it et al.} reported recently the combined 
measurement of the experimental lifetime of the hypernuclei $\isotope[3][\Lambda]{H}$ and $\isotope[4][\Lambda]{H}$~\cite{Rappold:2013fic,Rappold:2014jqa}, and found that their lifetimes are significantly shorter than the free $\Lambda$ lifetime. This result was then confirmed by the ALICE~\cite{Adam:2015yta} and STAR~\cite{star} collaborations and raises crucial questions concerning the hyperon decay mechanisms in nuclear environment, especially in light nuclei.

In this work, we revisit the free hyperon decay and confirm the dominance of the pole contributions via the baryon internal conversion process. Then, we show that there exists a strong cancellation between two pole terms which makes the lifetime of the free $\Lambda$ to be ``fine-tuned" to its present small value. In the case of the hadronic decays of light hypernuclei such as $\isotope[3][\Lambda]{H}$ and $\isotope[4][\Lambda]{H}$, we find that these two pole terms will be affected differently by the spectator nucleons. As a consequence, the fine-tuned cancellation in the free $\Lambda$ decays is broken and the transition amplitude is enhanced. It leads to a shortening of the lifetimes of $\isotope[3][\Lambda]{H}$ and $\isotope[4][\Lambda]{H}$ in their pionic weak decays.

\begin{figure}[!htc]
  \centering
  \includegraphics[width=0.5\textwidth]{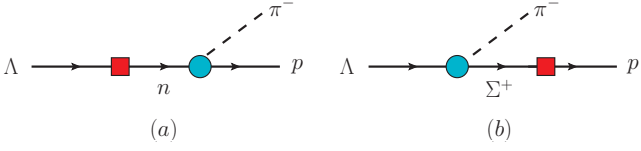}\\
  \caption{Feynman diagrams for the hadronic weak decay of a free $\Lambda\to p\pi^-$.}
  \label{fig-1}
\end{figure}

We concentrate on the baryon internal conversion process in this work since it is by orders of magnitude larger than the direct weak emission of pion in the hyperon decays. We start with the $\Lambda$ hadronic weak decay. The dominant diagrams for the baryon conversion process are shown in Fig.~\ref{fig-1}. The amplitude can be calculated in the quark model as:
\begin{multline}\label{eq-01}
{\cal M}= \langle p| H_\pi |n\rangle \frac{i}{\slashed{p} -m_n} \langle n|H_w^{PC} |\Lambda\rangle   \\
{}+ \langle p| H_w^{PC} |\Sigma^+ \rangle \frac{i}{\slashed{p} -m_\Sigma} \langle \Sigma^+|H_\pi |\Lambda\rangle \ ,
\end{multline}
where $H_\pi$ and $H_w^{PC}$ are Hamiltonians for the strong and parity conserved (PC) weak couplings and the transition matrix elements can be worked out in the framework of the constituent quark model (CQM). Given that the baryon wavefunctions are anti-symmetrized, the explicit expansion of the weak interaction Hamiltonian gives:
\begin{multline}\label{eq-02}
H_w^{PC}(1,2)=\frac{G_F}{\sqrt{2}} \cos\theta_C\sin\theta_C\langle \tilde{B}_f(1,2,3)|\tau_1^{(-)} v_2^{(+)} \\
{}\times (1-\vec{\sigma}_1.\vec{\sigma}_2)\delta(\vec{r}_1-\vec{r}_2)|\tilde{B}_i(1,2,3)\rangle
\ ,
\end{multline}
where $|\tilde{B}_i(1,2,3)\rangle$ and $|\tilde{B}_f(1,2,3)\rangle$ denote the internal quark wavefunctions for the initial- and final-state baryons, respectively; $\tau_1^{(-)}$ and $v_2^{(+)}$ are the flavor-changing operators that lower the isospin of quark number 1 and raise the strangeness of quark number 2, respectively. The leading weak transition operator does not flip the quark spins. The transition matrix element is sensitive to the short-distance structure of the quark wavefunctions due to the $\delta$ function in \eqref{eq-02}.  But for weakly bound nuclei or hypernuclei, the internal quark motion is not much modified as that for the free baryons. As the result, the uncertainties arising from the short-distance character should not change drastically for the decays of hyperons within the nuclear medium.

The  pion emission can be studied in the chiral quark model~\cite{Manohar:1983md,Li:1997gd,Zhao:2002id}, with a transition Hamiltonian
\begin{equation}\label{chiral-hamiltonian}
H_\pi = \frac{1}{f_\pi}\sum_j \bar{\psi}_j\gamma_\mu\gamma_5\partial^\mu \phi_\pi \hat{I}^\pi_j \psi_j \ ,
\end{equation}
where $f_\pi$ is the pion decay constant, $j$ denotes the $j^{\text {th}}$ quark in the baryons which interacts with the emitted pion and $\hat{I}^\pi_j$ is the corresponding flavor operator.

After the nonrelativistic expansion, the transition amplitude for $\Lambda\to p\pi^-$  can then be written as
\begin{equation}\label{amp-lambda-2}
{\cal M} = \hat{\cal V}\,   {\cal G}(\Lambda\to p\pi^-)\ ,
\end{equation}
where $\hat{\cal V}$ is a common function
\begin{multline}\label{v-func}
\hat{\cal V}  \equiv   \sqrt{2M_i(M_f+E_f)} \left(1+\frac{q_0}{E_f+M_f}+\frac{q_0}{3\mu_q}\right)\\
{}\times 12 \ |\vec{q}| \exp\left[-\frac{\vec q^2}{6\,\alpha_h^2}\right]\,\left(\frac{\alpha_h}{\sqrt\pi}\right)^3 G_F\cos\theta_C\sin\theta_C \ ,
\end{multline}
where $E_i$ and $E_f$ denote the energies carried by the baryons before and after emitting pion, $(q_0,\vec q)$ is the pion four-momentum in the c.m. frame, $\mu_q$ is the reduced mass of two interacting quarks and in our case, $\mu_q\simeq m_u/2\simeq m_d/2$, and the exponential factor  is due to the separation of the c.m.\ and internal motions of the quark system and extracted in the approximation of a simple harmonic-oscillator of strength $\alpha_h$. This factor plays the role of a form factor for the axial-vector coupling when the internal quark structure is considered.

In Eq.~(\ref{amp-lambda-2}), ${\cal G}$ is a channel-dependent function, i.e., for $\Lambda\to p\pi^-$,
\begin{multline}\label{amp-pi-p}
{\cal G}_{(\Lambda\to p\pi^-)}
\equiv \biggl[ \frac{ g_{np\pi^-}C^W_{(\Lambda\to n)} }{M_\Lambda^2-M_n^2}+\frac{g_{\Lambda\Sigma^+\pi^-}C^W_{(\Sigma^+\to p)}}{M_p^2-M_\Sigma^2}\biggr] \ ,
\end{multline}
where  $C^W$ is the spin-flavor factor of the baryon internal conversion, and can be explicitly calculated in the SU(6) quark model. Its values for different processes are listed in Tab.~\ref{tab-cw}.

\begin{table}[!htc]
\caption{Weak matrix element $C^W_{(A\to B)}\equiv \langle B|\hat{\cal O}^W|A\rangle$ for the baryon conversions, with  $\hat{\cal O}^W\equiv\tau_1^{(-)} v_2^{(+)}(1-\mathbf{\sigma}_1 \cdot \mathbf{\sigma}_2)$. }
\label{tab-cw}
$$
\begin{array}{ccc}
\hline\hline
 \quad\langle n|\hat{\cal O}^W|\Lambda\rangle\quad  & \quad \langle p|\hat{\cal O}^W|\Sigma^+\rangle\quad   &
 \quad\langle n|\hat{\cal O}^W|\Sigma^0\rangle \quad \\
 \hline\\[-8pt]
 -1/\sqrt{6}   &  +1   &   1/\sqrt{2} \\
\hline\hline
\end{array}
$$
\end{table}
The hadronic couplings of pseudoscalar mesons to the octet baryons are defined by the Goldberger-Treiman relation~\cite{Goldberger:1958tr}:
\begin{equation}\label{GT-relation}
g_{B_i B_f\pi}\equiv \frac{C_{B_i B_f\pi}\,g_A(B_i B_f\pi)\, \bar{M}}{f_\pi}  \ ,
\end{equation}
where $\bar{M}\equiv (M_i+M_f)/2$ is the averaged baryon mass of the interacting baryons. The departure from unity of parameter $C_{B_i B_f\pi}$  indicates a SU(3) flavor symmetry breaking.
The axial-vector coupling $g_A$  can be explicitly calculated in the chiral quark model via
\begin{equation}
g_A(B_i B_f\pi)\equiv \frac{\langle B_f|\sum_j \hat{I}^\pi_j \sigma_{jz} |B_i\rangle}{\langle B_f| \sigma^{tot}_z |B_i\rangle} \ ,
\end{equation}
where $\sigma_{jz}$ and $\sigma^{tot}_z$ are the quark and baryon spin operator projections to the $z$ axis, respectively. The values for $g_A$ in the SU(6) CQM are listed in Tab.~\ref{tab-axial}.

\begin{table}[!htbc]
\caption{Axial-vector couplings  for the pion emission.}
\label{tab-axial}
\begin{tabular}{lc@{\qquad}lc}
 \hline\hline
 Process \qquad      &  $g_A$ &  Process \qquad       & \quad $g_A$  \quad    \\
 \hline
 $p\to n\pi^+$ &   $5/3$ & $\Sigma^+\to\Lambda\pi^+$ &   $-2/\sqrt{6}$  \\
 $n\to p\pi^-$ &  $5/3$  & $\Sigma^-\to\Lambda\pi^-$ &   $-2/\sqrt{6}$  \\
 $n\to n\pi^0$ &   $5/(3\sqrt{2})$  & $\Sigma^+\to\Sigma^0\pi^+$ &  $4/(3\sqrt{2})$  \\
 $p\to p\pi^0$ &   $-5/(3\sqrt{2})$  & $\Sigma^+\to\Sigma^+\pi^0$ &  $-4/(3\sqrt{2})$ \\
 $\Lambda\to\Sigma^+\pi^-$ &  $-2/\sqrt{6}$  & $\Sigma^-\to\Sigma^0\pi^-$ &  $-4/(3\sqrt{2})$ \\
 $\Lambda\to\Sigma^0\pi^0$ &  $-2/\sqrt{6}$  &   &  \\
\hline\hline
\end{tabular}
\end{table}

The same analysis can be done for $\Lambda\to n\pi^0$, $\Sigma^+\to n \pi^+$ and $p\pi^0$, and $\Sigma^-\to n\pi^-$ which share similar dynamic mechanisms.
For $\Lambda\to n\pi^+$,  the difference arises from the strong pion emission vertices compared to Fig.~\ref{fig-1}. Therefore, one would expect that
$R\equiv {\Gamma(\Lambda\to p\pi^-)}/{\Gamma(\Lambda\to n\pi^0)}\simeq 2$, given the dominance of the baryon conversion processes in the $\Lambda$ hadronic decays. This relation actually agrees very well with  the experimental data. For $\Sigma^\pm$ pionic weak decays, the ${\cal G}$ functions have the following expressions:
\begin{eqnarray}
  {\cal G}_{(\Sigma^+\to n\pi^+)} &\equiv
\biggl[ \frac{ g_{pn\pi^+}C^W_{(\Sigma^+\to p)} }{M_\Sigma^2-M_p^2}+\frac{g_{\Sigma^+\Lambda\pi^+}C^W_{(\Lambda\to n)}}{M_n^2-M_\Lambda^2}\nonumber\\
& + \frac{g_{\Sigma^+\Sigma^0\pi^+} C^W_{(\Sigma^0\to n)}}{M_n^2-M_\Sigma^2}\biggr] \ ,\\
 {\cal G}_{(\Sigma^+\to p\pi^0)} &\equiv   C^W_{(\Sigma^+\to p)}\biggl[ \frac{ g_{pp\pi^0} }{M_\Sigma^2-M_p^2}+\frac{g_{\Sigma^+\Sigma^+\pi^0}}{M_p^2-M_\Sigma^2} \biggr] \ ,\\
{\cal G}_{(\Sigma^-\to n\pi^-)} &\equiv    \biggl[ \frac{g_{\Sigma^-\Lambda\pi^-}C^W_{(\Lambda\to n)}}{M_n^2-M_\Lambda^2}+ \frac{g_{\Sigma^-\Sigma^0\pi^-} C^W_{(\Sigma^0\to n)}}{M_n^2-M_\Sigma^2}\biggr] \ .\label{amp-lambda-3}
\end{eqnarray}

With the weak and strong couplings determined in the SU(3) flavor symmetry limit (see Tabs.~\ref{tab-cw} and \ref{tab-axial}), one recognizes that there exists an explicit cancellation among the pole terms for each process in Eqs.~(\ref{amp-pi-p}) and (\ref{amp-lambda-3}). As a consequence of such an intrinsic ``fine-tuned" cancellation the amplitudes for each process will be highly suppressed such that the lifetimes of these states are relatively long. Although the detailed cancellation will depend on models, there is no doubt that such intrinsic cancellations occur among the pole terms due to the SU(3) flavor symmetry. Thus, a natural prospect is that if the nuclear media act on those pole terms differently, they will break down the fine-tuned cancellation and result in significantly enhanced amplitudes. As follows, we will demonstrate that such a scenario indeed occurs.

It should be stressed that the relative signs determined by the SU(3) symmetry is essential for recognizing the underlying dynamics for the pionic weak decays. Meanwhile, a quantitative description of the data would require the inclusion of the SU(3) symmetry breaking effects. We explicitly adopt Eq.~(\ref{GT-relation}) in the fitting leaving $C_{B_i B_f\pi}$ to be fitted by experimental data. Namely, the values for $C_{B_i B_f\pi}$ deviating from unity will reflect the SU(3) symmetry breaking. We also treat $\alpha_h$ as a parameter to be fitted by experimental data. 
We find that with $\alpha_h=305.12\pm 0.75$ MeV, $C_{NN\pi}=0.843\pm 0.001$, $C_{\Lambda\Sigma\pi}=1.400\pm 0.086$, and $C_{\Sigma\Sigma\pi}=1.128\pm 0.002$, the experimental data can be reasonably described. Moreover, there exist strong correlations among $C_{B_i B_f\pi}$ and the SU(3) symmetry breaking is about $40\%$ at most. The fitted partial widths are listed in Tab.~\ref{tab-3} (3rd col.) to compare with the experimental data (4th col.)~\cite{Agashe:2014kda}. To demonstrate the sensitivity of the cancellation phenomena, we fix $\alpha_h=305.12$ MeV and $C_{B_i B_f\pi}=1$ (i.e. in the SU(3) symmetry limit) to extract the partial widths (2nd col.). 

\begin{table}[!htbc]
\caption{The partial decay widths for $\Lambda$ and $\Sigma^\pm$ pionic weak decays in unit of $10^{-6}$ eV. The second column is obtained in the SU(3) flavor symmetry limit. The third column is obtained by fitting parameters $C_{B_i B_f\pi}$. The experimental values are listed in the 4th column.  }
\label{tab-3}
\begin{tabular}{@{\qquad}c@{\qquad}c@{\qquad}c@{\qquad}c}
\hline\hline
Channels & SU(3)  & Fitting & Experimental data \\
\hline
$\Lambda\to p\pi^-$ & $0.65$  & $1.62^{+0.50}_{-0.43}$ & $1.60\pm 0.02 $  \\
$\Lambda\to n\pi^0$  & $0.35$  &  $0.91^{+0.28}_{-0.24}$ & $0.895\pm 0.014 $ \\
$\Sigma^+\to p\pi^0$ & $57.32$  &  $5.64^{+0.17}_{-0.17}$ & $4.23\pm 0.03$  \\
$\Sigma^+\to n\pi^+$ & $31.22$  &  $2.34^{+1.05}_{-0.85}$ & $3.96\pm 0.03$  \\
$\Sigma^-\to n\pi^-$ & $3.87$  &  $3.38^{+1.13}_{-0.97}$ & $4.44\pm 0.03 $  \\
\hline\hline
\end{tabular}
\end{table}

\begin{figure}[!htbc]
  \centering
  \includegraphics[width=0.5\textwidth]{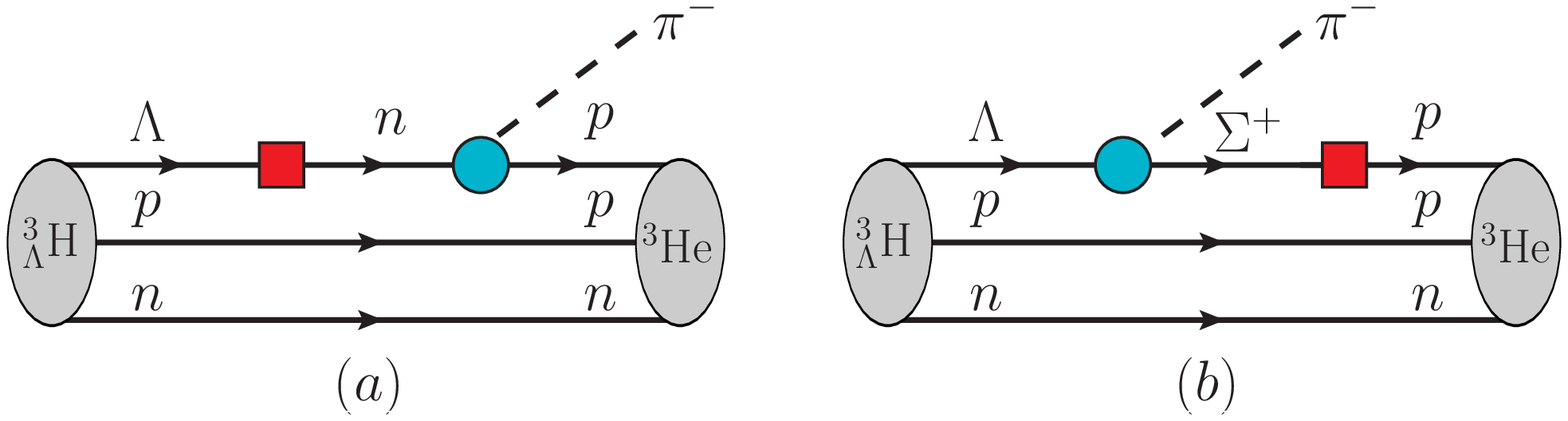}\\
  \caption{Feynman diagrams for $\isotope[3][\Lambda]{H}\to \isotope[3][]{He} + \pi^-$.}
  \label{fig-6}
\end{figure}

Proceeding to investigate the nuclear effects on the above scenario, we apply the harmonic oscillator wavefunctions for $\isotope[3][\Lambda]{H}$ and $\isotope[3][]{He}$ in momentum space:
\begin{align}
\Psi(\vec{p}_i)&=\int \tilde{\Psi}(\vec{r}_i) \delta^3(\vec{R})\Pi_i [\exp(-i \vec{p}_i\cdot\vec{r}_i) d^3\vec{r}_i]\\
&\nonumber =\frac{(\sum_i m_i)^3 N}{\Delta^{\frac 32}} \exp-\genfrac{[}{]}{}{0}{\sum_{i\neq j\neq k}\beta_i (m_j \vec{p}_k-m_k \vec{p}_j)^2}{2\Delta}\;,
\end{align}
with the normalization $\int \Psi(\vec{p}_i)^2\delta^3(\vec{P}) \prod_{i=1}^3 d^3\vec{p}_i= 1$ and the c.m.\ momentum $\vec{P}\equiv \sum_i \vec{p}_i$. Its Fourier transformed wavefunction in coordinate space is a simple  harmonic-oscillator,
$\tilde{\Psi}(\vec{r}_i)=N\exp[-\frac 12 \sum_i \beta_i r_i^2]$,
with $\vec{R}=\sum_i m_i\vec{r}_i/\sum_i m_i=0$ and normalization factor $N^2\equiv \pi^{-3}\Delta^{\frac 32} (m_1+m_2+m_3)^{-3}$ with  $\Delta\equiv m_3^2\beta_1\beta_2+m_2^2\beta_1\beta_3+m_1^2\beta_2\beta_3$.
The r.m.s.\ radii within $\tilde{\Psi}$ are \cite{coffou:1969}
\begin{equation}\label{eq-rms-beta}
\langle \vec{r}_i^2\rangle = \frac 32 \frac{m_j^2\beta_k+m_k^2\beta_j}{\Delta} \ ,
\end{equation}
with the  indices $(i,j,k)$ a permutation of $(1,2,3)$.
To determine the parameters $\beta_i$, we proceed in two steps. First, as in our previous work \cite{Richard:2014pwa}, the three-baryon problem is solved accurately, but with a simple monotonic potential for each pair, which reproduces the low-energy parameters. This leads to the radii shown in  Tab.~\ref{tab-beta}. Next, the $\beta_i$ are computed by solving Eq.~\eqref{eq-rms-beta} and their values are also listed in Tab.~\ref{tab-beta} for both the J\"ulich~\cite{Polinder:2007mp, Haidenbauer:2013oca} and Nijmegen \cite{Nijmegen} models. We note that the lack of hard core leads to an underestimate of the r.m.s.\ radii. In particular, the values for $\isotope[3]{He}$ in Tab.~\ref{tab-beta} are smaller than those determined by electron scattering and fitted by models~\cite{Sick:2015spa,Coon:2014nja}. Since the same strategy is applied to the wavefunctions for both strange and non-strange nuclei, we argue that the wavefunction overlap of the two $S$-wave ground states $\isotope[3][\Lambda]{H}$ and $\isotope[3][]{He}$ do not suffer much from this approximation.  In future refinements of this work, more realistic wavefunctions can be applied, such as
superpositions of Gaussians with different $\beta_i$ parameters in each term.

In the impulse approximation, the hadronic weak decay of $\isotope[3][\Lambda]{H}\to \isotope[3][]{He} + \pi^-$ can be regarded as due to the $\Lambda$ decaying into $p\pi^-$ while the initial proton and neutron remain as spectators. The transition can also occur via two processes as shown in Fig.~\ref{fig-6} $(a)$ and $(b)$ of which the corresponding elementary processes are described in Fig.~\ref{fig-1}. The transition amplitude can then be written in terms of operators at hadron level and the intermediate three-body propagators can be included. By expanding the propagator in a nonrelativistic form and integrating the energy part with the on-shell condition, the nuclear transition amplitude can be written as
\begin{multline}\label{nucl-trans}
{\cal M}= \int \frac{d {\bf p}_1}{(2\pi)^3} \frac{d {\bf p}_2}{(2\pi)^3} \frac{d {\bf p}_3}{(2\pi)^3} \Psi_{\isotope[3][]{He}}^*({\bf P}_f; {\bf p}_1, {\bf p}_2, {\bf p}_3-{\bf q})\\
\times  \frac{(2\pi i)^2 \langle \isotope[3][]{He}| H_\pi^{(3)}|[p, n, n]^a\rangle\langle [p, n, n]^a  | H_w^{(3)}| \isotope[3][\Lambda]{H}\rangle }{M_{\isotope[3][\Lambda]{H}}-(M_1+M_2+M_n)-(\frac{{\bf p}_1^2}{2M_1}+\frac{{\bf p}_2^2}{2M_2}+\frac{{\bf p}_3^2}{2M_n})} \\
\times \Psi_{\isotope[3][\Lambda]{H}}({\bf P}_i; {\bf p}_1, {\bf p}_2, {\bf p}_3) \delta({\bf p}_1+{\bf p}_2+{\bf p}_3-{\bf P}_i)\\
+  \int \frac{d {\bf p}_1^\prime}{(2\pi)^3} \frac{d {\bf p}_2^\prime}{(2\pi)^3} \frac{d {\bf p}_3^\prime}{(2\pi)^3} \Psi_{\isotope[3][]{He}}^*({\bf P}_f; {\bf p}_1^\prime, {\bf p}_2^\prime, {\bf p}_3^\prime)  \\
\times \frac{ (2\pi i)^2 \langle \isotope[3][]{He}| H_w^{(3)}|[p, n, \Sigma^+]\rangle\langle [p, n, \Sigma^+] | H_\pi^{(3)}| \isotope[3][\Lambda]{H}\rangle}{E_{\isotope[3][]{He}}-(M_1+M_2+M_\Sigma)-(\frac{{\bf p}_1^{\prime 2}}{2M_1}+\frac{{\bf p}_2^{\prime 2}}{2M_2}+\frac{{\bf p}_3^{\prime 2}}{2M_\Sigma})}\\
\times  \Psi_{\isotope[3][\Lambda]{H}}({\bf P}_i; {\bf p}_1^\prime, {\bf p}_2^\prime, {\bf p}_3^\prime+{\bf q}) \delta({\bf p}_1^\prime+{\bf p}_2^\prime+{\bf p}_3^\prime-{\bf P}_f) \ ,
\end{multline}
where the initial  $\isotope[3][\Lambda]{H}$ and final $\isotope[3][]{He}$ contain only $S$-wave components as the leading approximation and they are both anti-symmetrized in the spin-isospin space in order to respect the Fermi statistics.

\begin{table}[!htdp]
\caption{Parameters $\beta_i$ extracted by fitting the r.m.s. radii from J\"ulich (I)~\cite{Polinder:2007mp,Haidenbauer:2013oca} and Nijmegen (II) model~\cite{Nijmegen}.}
\begin{center}
\begin{tabular}{|c|c|c|c|c|c|c|}
\hline\hline
System & $r_n (\mathrm{fm})$ & $r_p  (\mathrm{fm})$ & $r_\Lambda  (\mathrm{fm})$& $\beta_n  (\mathrm{fm}^{-2})$ & $\beta_p (\mathrm{fm}^{-2})$ & $\beta_\Lambda (\mathrm{fm}^{-2})$\\\hline
$\isotope[3][]{He} $ & 1.38 & 1.49 &--& 0.430 & 0.573 &--\\
$\isotope[3][\Lambda]{H}$ (I) & 1.60 & 1.60 & 1.65 & 0.469 & 0.469 & 0.220\\
$\isotope[3][\Lambda]{H}$ (II) & 2.32 & 2.32 & 2.84 & 0.296 & 0.296 & -0.023\\
\hline\hline
\end{tabular}
\end{center}
\label{tab-beta}
\end{table}

In Eq.~(\ref{nucl-trans}) the notation $|[p,n,n]^a\rangle$ denotes the anti-symmetrization requirement on the intermediate $[p,n,n]$ system due to Fermi statistics, while the intermediate $|[p,n,\Sigma^+]\rangle$ does not have such a constraint. Compared to Fig.~\ref{fig-1} (a) of the free $\Lambda$ decays the intermediate neutron of Fig.~\ref{fig-6} (a) will be affected by the spectator neutron which has taken half of the ground spin states for the $[p,n,n]$ system if we neglect the virtual effects. This will break the ``fine-tuning" of cancellation between those two pole terms in Fig.~\ref{fig-1}. Interestingly, there exists a kinematic effect to compete against the Pauli blocking in Fig.~\ref{fig-6} (a). Since the mass of nucleon is smaller than $\Lambda$ it allows the intermediate $[p,n,n]$ to be on shell in certain kinematic region. It corresponds to a three-body pole structure in the transition matrix element and will enhance the amplitude of Fig.~\ref{fig-6} (a) which again will violate the fine-tuned
cancellation in the free $\Lambda$ decays.

Such effects can be examined by explicit calculations adopting the parameterized wavefunctions for  $\isotope[3][\Lambda]{H}$ and $\isotope[3][]{He}$. Although this is a crude approximation its effects generally will decrease the amplitude compared to the case of free $\Lambda$ due to the wavefunction convolution. In other words, one should not expect a significant enhancement of the amplitude caused by the nuclear wavefunctions. In Tab.~\ref{part-wid}, the calculated partial width $\Gamma(\isotope[3][\Lambda]{H}\to \isotope[3][]{He} +\pi^-)=2.18\times 10^{-6}$ eV, is listed. It is larger than that for free $\Lambda$, i.e. $(1.60\pm 0.02)\times 10^{-6}$ eV~\cite{Agashe:2014kda}. Meanwhile, significant cancellations between  Fig.~\ref{fig-6} (a) and (b) can be seen by comparing their exclusive contributions to the full result.

\begin{figure}[!htbc]
  \centering
  \includegraphics[width=0.4\textwidth]{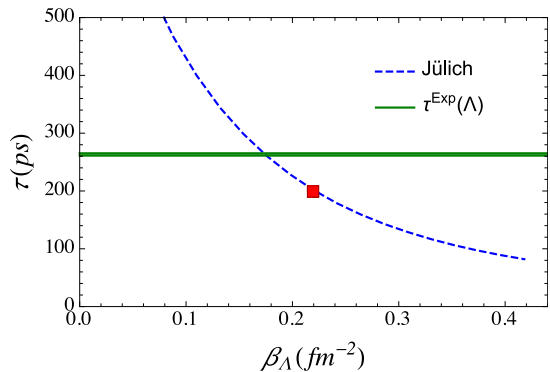}\\
  \vspace{0cm}
  \caption{Sensitivity of the lifetime to the nuclear wavefunctions studied by varying $\beta_\Lambda$, but keep the other two $\beta$ values   fixed in the J\"ulich model for $\isotope[3][\Lambda]{H}$. The red square corresponds to the lifetime with the $\beta_\Lambda$ extracted from the r.m.s. radii, while the horizontal band is the experimental data for the free $\Lambda$. }
  \label{fig-beta-depend}
\end{figure}

Being aware of that the partial width of $\isotope[3][\Lambda]{H}\to \isotope[3][]{H} +\pi^0$ is just half of  $\isotope[3][\Lambda]{H}\to \isotope[3][]{He} +\pi^-$ in the spin-flavor symmetry limit, we obtain the two-body decay width of $3.27\times 10^{-6}$ eV. Neglecting the contributions from other possible channels, such as  $\isotope[3][\Lambda]{H}\to d + p +\pi^-$, $d+ n +\pi^0$, $p+p+n+\pi^-$, and $p+n+n+\pi^0$, the estimated lifetime is $\tau(\isotope[3][\Lambda]{H})\simeq 2.0 \times 10^{-10}~\mathrm{s}$, which is significantly shorter than the one of the free $\Lambda$,
$\tau(\Lambda)=(2.63\pm 0.020)\times 10^{-10}~\mathrm{s}$~\cite{Agashe:2014kda}.
In Tab.~\ref{lifetime}, this calculated value is compared with the most recent experimental measurements and sensitivities of the proposed mechanism to the nuclear wavefunctions is shown in Fig.~\ref{fig-beta-depend} by varying $\beta_\Lambda$ but with the other two $\beta$ values fixed in the J\"ulich model. We emphasize that we demonstrate the essential reaction mechanism instead of fully quantify it. Therefore, although there should be uncertainties with the estimated partial width for  $\isotope[3][\Lambda]{H}\to \isotope[3][]{He} +\pi^-$, it is clear that the nuclear effects which violate the fine-tuned cancellation will result in enhanced amplitudes. This should be the key for understanding the recently observed shortened lifetime for $\isotope[3][\Lambda]{H}$. In particular, note that such a mechanism can arise from light nucleus system instead of heavy ones.

There is an abundant literature on the weak decay of hypernuclei \cite{Alberico:2001jb}, in particular dealing with the importance of pionless decays $\Lambda n\to n n$ and $\Lambda p \to p p$. In the case of hypertriton, Kamada et al.\ \cite{Kamada:1997rv}  studied sophisticated final-state corrections, assuming a frozen vertex for the weak decay, and found a very small departure from the case of free $\Lambda$.  Our study of the weak decay provides a novel mechanism to explain the shortening of the hypertriton lifetime.

\begin{table}[!htdp]
\caption{The partial width of $\isotope[3][\Lambda]{H}\to \isotope[3][]{He} +\pi^-$ calculated with parameters fitted by the J\"ulich model~\cite{Polinder:2007mp, Haidenbauer:2013oca}. Contributions from Fig.~\ref{fig-6} (a) and (b),  and the sum of both are listed individually. The cancellation in $\isotope[3][\Lambda]{H}$ decay is not as strong as that in $\Lambda$ decay due to the wavefunction convolution of the initial and final nuclei. We do not show results from the Nijmegen model since the harmonic oscillator interpretation does not work as indicated by the negative $\beta_{\Lambda}$ value.}
\begin{center}
\begin{tabular}{|c|c|c|c|}
\hline\hline
$\Gamma(\isotope[3][\Lambda]{H}\to \isotope[3][]{He} +\pi^-) (10^{-6} \mathrm{eV})$ & (a) & (b) &Total\\\hline
J\"ulich model & 3.25 & 10.75 & 2.18 \\\hline\hline
\end{tabular}
\end{center}
\label{part-wid}
\end{table}%

\begin{table}[htdp]
\caption{Recently measured lifetimes of $\isotope[3][\Lambda]{H}$ (in units of $ps$) compared with our theory result with which the error is calculated by varying $\beta_\Lambda$ within $10\%$. This is to show the sensitivity of the cancellation mechanism to the nuclear wavefunctions instead of uncertainty estimate. }
\begin{center}
\begin{tabular}{|c|c|c|c|c|}
\hline\hline
Ref.~\cite{Rappold:2014jqa}  & Ref.~\cite{Rappold:2013fic} & Ref.~\cite{Adam:2015yta} &Ref.~\cite{star} & Theory\\\hline
$217^{+19}_{-16}$ & $183^{+42}_{-32}\pm 37$ & $181^{+54}_{-39}\pm 33 $& $155^{+25}_{-22}\pm 29$ & $200\pm 23 $ \\
\hline
\end{tabular}
\end{center}
\label{lifetime}
\end{table}%

Useful discussions with C. Hanhart, Ulf.-G Mei{\ss}ner and A. Nogga, as well as correspondence with Jin-Hui Chen, Yu-Gang Ma, Zhangbu Xu and Sidney Coon are greatly appreciated. This collaboration was made possible by the China--France program FCPPL. This work is also supported, in part,
by the National Natural Science Foundation of China (Grant Nos.
11035006 and 11425525), and the Sino-German CRC 110 ``Symmetries and
the Emergence of Structure in QCD" (NSFC Grant No. 11261130311).

\end{document}